\documentstyle[12pt]{article}

\begin{document}
\parskip 1ex
\setcounter{page}{1}
\oddsidemargin 0pt
 %   Note that \oddsidemargin =  \evensidemargin
\evensidemargin 0pt
\topmargin -40pt
 %    Nominal distance from top of page to  top of \jot = .5ex
%
%%%%%%%%%%%%%%%%%%%%%%%%%%%%%%%%
\newcommand{\be}{\begin{equation}}
\newcommand{\ee}{\end{equation}}
\newcommand{\beq}{\begin{eqnarray}}
\newcommand{\eeq}{\end{eqnarray}}
\def\a{\alpha}
\def\b{\beta}
\def\g{\gamma}
\def\G{\Gamma}
\def\d{\delta}
\def\e{\epsilon}
\def\z{\zeta}
\def\h{\eta}
\def\th{\theta}
\def\k{\kappa}
\def\l{\lambda}
\def\L{\Lambda}
\def\m{\mu}
\def\n{\nu}
\def\x{\xi}
\def\X{\Xi}
\def\p{\pi}
\def\P{\Pi}
\def\r{\rho}
\def\s{\sigma}
\def\S{\Sigma}
\def\t{\tau}
\def\f{\phi}
\def\F{\Phi}
\def\c{\chi}
\def\w{\omega}
\def\W{\Omega}
\def\de{\partial}

%%%%%%%%%%%%%%%%%%%%%%%%%%%%%%%%%%%%
%Without pictures use this macro
\def\pct#1{(see Fig. #1.)}
%%%%%%%%%%%%%%%%%%%%%%%%%%%%%%%%%%%
%With pictures use this macro
%\def\pct#1{\input epsf \centerline{ \epsfbox{#1.eps}}}

%%%%%%%%%%%%%%%%%%%%%%%% FRONT PAGE %%%%%%%%%%%%%%%%%%%%%%%%%%%%%%%%%%%%%
\begin{titlepage}
\hbox{\hskip 12cm ROM2F-99/41  \hfil}
\hbox{\hskip 12cm hep-th/9910246 \hfil}
\hbox{\hskip 12 cm \today}
%\end{flushright}
\vskip 1.4cm
\begin{center} 
{\Large  \bf  Abelian \ Vectors  \ and \ Self-Dual \ Tensors \vskip .6cm
in \ Six-Dimensional \ Supergravity}

\vspace{1.8cm}
 
{\large \large Fabio Riccioni}
\vspace{0.6cm}

{\sl Dipartimento di Fisica, \ \
Universit{\`a} di Roma \ ``Tor Vergata'' \\
I.N.F.N.\ - \ Sezione di Roma \ ``Tor Vergata'', \\
Via della Ricerca Scientifica , 1 \ \ \
00133 \ Roma \ \ ITALY}
\end{center}
\vskip 1.5cm

\abstract{In this note we describe the most general coupling of
{\it abelian} vector and tensor multiplets to six-dimensional
$(1,0)$ supergravity. As was recently pointed out, it is of interest 
to consider more general Chern-Simons couplings to abelian vectors
of the type $H^{r}=d 
B^{r}-1/2 \ c^{rab} A^{a}d A^{b}$, with $c^{r}$ matrices that may not be 
simultaneously diagonalized. We show that these couplings can be 
related to Green-Schwarz terms of the form
$B^r c_r^{ab} F^a F^b$, and how 
the complete local Lagrangian, that embodies 
factorized gauge and supersymmetry anomalies
(to be disposed of by fermion loops) is uniquely determined by 
Wess-Zumino consistency conditions, aside from an arbitrary quartic 
coupling for the gauginos.}
\vskip 36pt
\hbox{\hskip 1.2cm Revised version: November 2000 \hfil}
\vfill
\end{titlepage}
%%%%%%%%%%%%%%%%%%%%%%%%%%%%%%%%%%%%%%%%%%%%%%%%%%%%
\makeatletter
\@addtoreset{equation}{section}
\makeatother
\renewcommand{\theequation}{\thesection.\arabic{equation}}
\addtolength{\baselineskip}{0.3\baselineskip} 
%%%%%%%%%%%%%%%%%%%%%%%%%%%%%%%%%%%%%%%%%%%%%%%%%%%%

Vector multiplets coupled to variable numbers of tensor multiplets
in six-dimensional $(1,0)$ supergravity 
arise naturally in perturbative
type-I vacua \cite{bs}, that are related by string dualities to 
non-perturbative heterotic and M-theory vacua. 
The low-energy field equations of these $(1,0)$ models
have revealed the explicit realization of a peculiar aspect of the 
physics of branes: singularities in the gauge couplings appear 
for particular values of the scalars in the tensor multiplets 
\cite{as}, and  
can be ascribed to phase transitions \cite{dmw} in which a
string becomes tensionless \cite{tensionless}. 

In these models, that arise as parameter-space orbifolds
(orientifolds) \cite{cargese}
of $K3$ reductions of the type-IIB string, several antisymmetric
tensors take part in a generalized
Green-Schwarz mechanism \cite{gs,as}, and the resulting residual anomaly
polynomial has the form 
$$
\ c^r_x \ c^s_y \ \eta_{rs} \ {\rm tr}_x F^2 \ {\rm tr}_y F^2
\quad ,
$$
where the $F$'s denote collectively the Riemann curvature and the 
Yang-Mills field strengths,
the $c$'s are a collection of constants ($x$ and $y$ run over the various
semi-simple Lie factors in the gauge group and over the Lorentz group) 
and $\eta$ is the Minkowski metric for
$SO(1,n)$, with $n$ the number of tensor multiplets.
Taking into account only the gauge part of the anomaly polynomial,
the corresponding Green-Schwarz term, 
$$
B^r \ c_r^x \ {\rm tr}_x F^2 \quad , 
$$
contains two derivatives, and thus belongs to the low-energy effective action.
Consequently, the Lagrangian has a ``classical'' gauge anomaly, that the 
Wess-Zumino  conditions \cite{wz} relate to a ``classical'' supersymmetry
anomaly \cite{fms}.
This model is thus rather different from
the other supergravities, since it is naturally determined by 
Wess-Zumino  conditions,
rather than by the usual requirement of local supersymmetry. 
As a consequence, the supersymmetry algebra contains a two-cocycle, 
and the resulting Lagrangian is determined up to an arbitrary term proportional
to the square of a bilinear in the gauginos \cite{frs}.
Moreover, the divergence of the energy-momentum 
tensor is non-vanishing \cite{rs}, as is properly the case for a theory that 
has  gauge anomalies but no gravitational anomalies, that 
could be accounted for introducing  higher-derivative couplings.

The coupling of $(1,0)$ supergravity
to $n$ tensor multiplets was originally studied
in \cite{romans} to lowest order in the fermi 
fields, while \cite{ns1} considered the coupling 
to a single tensor multiplet and to vector and hyper-multiplets 
to all orders in the fermi fields.  
In this case the kinetic term is proportional to $a 
e^{\phi}+b e^{-\phi}$ \cite{as}, where $a$ and $b$ are constants and $\phi$ is 
the scalar in the tensor multiplet, while \cite{ns1} actually deals with 
the particular case $a=0$, in which the anomaly 
polynomial vanishes and no tensionless string transition occurs. 
The general coupling to non-abelian vectors and self-dual tensors
was worked out to lowest order in the 
fermi fields in \cite{as}. In this covariant formulation the 
requirement of supersymmetry gives non-integrable equations, and the 
divergence of the vector equation gives the covariant anomaly.
The same model was then reconsidered again to lowest order in the fermi fields
in \cite{fms} in the consistent formulation, requiring the closure of the
Wess-Zumino conditions, that relate the consistent gauge anomaly to the
supersymmetry anomaly.
Additional couplings, as well as the 
inclusion of hyper-multiplets, were then considered in \cite{ns2}.
The complete coupling to non-abelian vector and tensor multiplets 
was finally obtained in \cite{frs} in the 
consistent formulation and in \cite{rs} in the covariant formulation. 

One can actually consider a slight modification of these couplings, 
resulting from the inclusion in the low-energy Lagrangian of Green-Schwarz 
terms non-diagonal in a set of $U(1)$ gauge groups.
A direct indication of how this can work is provided 
by a recent paper of Cremmer, Julia,
L\"u and Pope \cite{cjlp} where, in the search of the highest-dimensional 
origin of various three-dimensional scalar sigma models, 
it is shown that the ``oxidation endpoint'' of 
three-dimensional supergravity with scalars in $F_4$ sigma models is  
$(1,0)$ six-dimensional supergravity coupled to two tensor multiplets
and two abelian vector multiplets, with non-diagonal Chern-Simons 
couplings. The corresponding kinetic terms are also 
non-diagonal, compatibly with the abelian gauge invariances.
The resulting model actually motivates the study of more
general Green-Schwarz couplings to {\it abelian} 
vectors of the form
$$
B^r \ c_r^{ab} \ F^a \ F^b \quad , 
$$
where the indices $a,b$ run over the different $U(1)$ gauge groups, while the 
symmetric matrices
$c^{r}$ may not be 
simultaneously diagonalized.
In this paper we describe this setting in detail,
constructing the general coupling of $(1,0)$ six-dimensional supergravity 
to $n$ tensor multiplets and {\it abelian} vector multiplets, and 
then pointing out the 
connection with the particular case considered in \cite{cjlp}. 

The $n$ scalars in the tensor 
multiplets parameterize the coset space $SO(1,n)/SO(n)$, and are described
by the $SO(1,n)$ matrix \cite{romans}
$$
V =\pmatrix{v_r \cr x^m_r}\quad .
$$
All spinors are symplectic Majorana-Weyl, the 
tensorinos $\chi^m$ $(m=1,\ldots,n)$
being right-handed and the gravitino $\psi_\m$ and the gauginos $\l$
being left-handed (we follow the notation of \cite{frs}). 
The tensor fields $B^r_{\m\n}$ are valued in the fundamental 
representation of $SO(1,n)$, and their field strengths include 
generalized
Chern-Simons 3-forms of the vector fields \cite{cjlp} according to
$$
H^r_{\m\n\r}  = 3 \partial_{[\m} B^r_{\n\r ]} - 3 c^{rab} A_{[\m}^{a}
\partial_{\n}A_{\r ] }^{b}\quad ,
$$
where the $c^{rab}$ are constants that determine the gauge part of the
residual anomaly polynomial
$$
c_{r}^{ab} c^{rcd} F^{a} \wedge F^{b} \wedge F^{c}
\wedge F^{d}\quad .
$$
In the complete theory, the anomaly induced by this term would 
cancel against the contribution of fermion loops, 
while the irreducible part of the anomaly polynomial is directly 
absent in consistent models \cite{rdsss,as}. 
The gauge invariance of $H^r$ requires that 
$$
\delta B^r =\frac{1}{2} c^{rab} \L^{a} dA^{b} \quad .
$$
The tensor fields satisfy (anti)self-duality conditions, conveniently
summarized as \cite{fms,frs}
$$
G_{rs} \hat{\cal{H}}^{s \m\n\r} =\frac{1}{6e} \e^{\m\n\r\a\b\g} 
\hat{\cal{H}}_{r \a\b\g}\quad ,
\label{selfdual}
$$
where $G_{rs} =v_r v_s + x^m_r x^m_s$ and
$$
\hat{\cal{H}}^{r}_{\m\n\r}=\hat{H}^{r}_{\m\n\r}-\frac{i}{8}
v^{r}(\bar{\chi}^{m}\g_{\m\n\r}\chi^{m})+\frac{i}{8}c^{rab}(\bar{\l}^{a}
\g_{\m\n\r}\l^{b})\quad ,
$$
with $\hat{H}$ the supercovariantization of $H$.
The model can be constructed using the method of \cite{frs},
where the complete field equations for the case of arbitrary numbers
of tensor and {\it non-abelian} vector multiplets were 
obtained requiring the closure of Wess-Zumino consistency conditions.
The completion to all orders in the fermi fields of the equations of motion
is obtained requiring the closure of the commutator of two supersymmetry 
transformations on the fermionic field equations. 
All the resulting equations may be conveniently derived from the Lagrangian
\beq  
e^{-1}{\cal{L}} & & =-\frac{1}{4}R +\frac{1}{12}G_{rs} H^{r \m\n\r}
H^s_{\m\n\r} -\frac{1}{4} \de_\m v^r \de^\m v_r-\frac{1}{4} v_r c^{rab} 
F^{a}_{\m\n} F^{b\m\n} 
\nonumber\\ & & -\frac{1}{16e}
\e^{\m\n\a\b\g\delta} c_r^{ab} B^r_{\m\n} F^{a}_{\a\b} F^{b}_{\g\delta}  
-\frac{i}{2}\bar{\psi}_\m \g^{\m\n\r} D_\n [\frac{1}{2}(\w +\hat{\w} )]
\psi_\r \nonumber \\
& & 
-\frac{i}{8}v_r [H+\hat{H}]^{r \m\n\r}(\bar{\psi}_\m \g_\n \psi_\r)
+\frac{i}{48} v_r [H+\hat{H} ]^r_{\a\b\g} (\bar{\psi}_\m
\g^{\m\n\a\b\g}\psi_\n ) \nonumber \\ & &
+\frac{i}{2} \bar{\chi}^m \g^\m D_\m (\hat{\w})
\chi^m  -\frac{i}{24}v_r \hat{H}^r_{\m\n\r} (\bar{\chi}^m
\g^{\m\n\r}
\chi^m ) \nonumber \\ & &
+\frac{1}{4}x^m_r [\de_\n v^r +\hat{\de_\n v^r} ](\bar{\psi}_\m \g^\n
\g^\m \chi^m) -\frac{1}{8} x^m_r [H+\hat{H}]^{r \m\n\r} (
\bar{\psi}_\m
\g_{\n\r}
\chi^m )\nonumber \\ & &
+\frac{1}{24}x^m_r [H+\hat{H}]^{r \m\n\r} (\bar{\psi}^\a \g_{\a\m\n\r}
\chi^m ) 
+\frac{1}{8}(\bar{\chi}^m \g^{\m\n\r} \chi^m )(\bar{\psi}_\m
\g_\n \psi_\r )\nonumber \\ & &
-\frac{1}{8}(\bar{\chi}^m \g^\m \chi^n )(\bar{\chi}^m \g_\m
\chi^n ) +
\frac{i}{4\sqrt{2}}  v_r c^{rab} (F+\hat{F})^{a}_{\n\r}(\bar{\psi}_\m
\g^{\n\r}\g^\m \l^{b} ) \nonumber \\ & &
 +\frac{1}{2\sqrt{2}}x^m_r c^{rab}(\bar{\chi}^m \g^{\m\n}
\l^{a} )\hat{F}^{b}_{\m\n}  +\frac{i}{2} v_r c^{rab}  (\bar{\l}^{a} \g^\m
\hat{D}_\m \l^{b} )\nonumber \\ & &
+\frac{i}{24} x^m_r x^m_s \hat{H}^r_{\m\n\r} c^{sab} 
(\bar{\l}^{a}\g^{\m\n\r} \l^{b} )
+\frac{1}{32}v_r c^{rab} (\bar{\l}^{a} \g_{\m\n\r} 
\l^{b}
)(\bar{\chi}^m
\g^{\m\n\r} \chi^m ) \nonumber\\  & & - \frac{i}{16}(\bar{\chi}^m
\g_{\m\n}\psi_\r )x^m_r c^{rab}  (\bar{\l}^{a} \g^{\m\n\r} \l^{b} ) 
 - \frac{i}{4} x^m_r c^{rab} (\bar{\chi}^m \g^\m \g^\n \l^{a} ) (\bar{\psi}_\m
\g_\n \l^{b})
\nonumber\\ & & -\frac{1}{16} v_r c^{rab}(\bar{\chi}^m \l^{a} )
(\bar{\chi}^m \l^{b})- \frac{3}{32}v_r c^{rab}(\bar{\chi}^m \g_{\m\n} 
\l^{a} )
(\bar{\chi}^m
\g^{\m\n}
\l^{b} ) 
\nonumber\\ & & 
+[(x^m \cdot c )(v \cdot c )^{-1} (x^n \cdot c )]^{ab} \lbrace
-\frac{1}{4}(\bar{\chi}^m \l^a )(\bar{\chi}^n \l^b )+
\frac{1}{16} (\bar{\chi}^n \g_{\m\n} \l^a )(\bar{\chi}^m \g^{\m\n} \l^b )
\nonumber\\
& & -\frac{1}{8} (\bar{\chi}^n \l^a )(\bar{\chi}^m \l^b )\rbrace
+\frac{1}{8} v_{r}c^{rab}(\bar{\psi}_\m \g_\n \psi_\r 
)(\bar{\l}^{a} \g^{\m\n\r}
\l^{b} ) 
\nonumber \\
& & 
 -\frac{1}{8} v_r v_s c^{rab}c^{s cd} (\bar{\l}^{a}\g_\m
\l^c )(\bar{\l}^{b} \g^\m \l^d ) 
+ \frac{\a}{8}c^{rab} c_r^{cd} (\bar{\l}^{a}
\g_\m
\l^c )(\bar{\l}^{b} \g^\m \l^d )
\quad,
\nonumber 
\eeq 
after imposing the (anti)self duality conditions. 
The last term, proportional to the arbitrary parameter $\a$, 
vanishes identically in the case of a single 
abelian vector multiplet. Since the kinetic terms of the vector fields 
are non-diagonal, this generalization is only possible 
in the abelian case. 

The variation of this Lagrangian with respect to gauge transformations 
gives the gauge anomaly
$$
{\cal{A}}_{\L}=-\frac{1}{32}\e^{\m\n\a\b\g\d}c_{r}^{ab}c^{rcd}\L^{a}F^{b}_{\m\n}
F^{c}_{\a\b}  F^{d}_{\g\d}\quad ,   
$$
while the variation with respect to the supersymmetry transformations
\beq 
& & \delta e_\m{}^\a =-i(\bar{\e} \g^\a \psi_\m ) \quad,\nonumber\\  
& & \delta B^r_{\m\n} =i v^r (\bar{\psi}_{[\m} \g_{\n]} \e )+ \frac{1}{2} x^{mr}
(\bar{\chi}^m
\g_{\m\n} \e )- c^{rab}  (A^{a}_{[\m}\delta A^{b}_{\n]}) \quad,
\nonumber\\  
& & \delta v_r = x^m_r (\bar{\chi}^m \e )\quad,\nonumber\\  & &
\delta A^{a}_\m = -\frac{i}{\sqrt{2}} (\bar{\e} \g_\m \l^{a} ) \quad ,
\nonumber\\  & &
\delta \psi_\m =\hat{D}_\m \e +\frac{1}{4} v_r \hat{H}^r_{\m\n\r}
\g^{\n\r}\e -\frac{3i}{8} \g_\m \chi^m (\bar{\e} \chi^m ) -\frac{i}{8} \g^\n
\chi^m (\bar{\e} \g_{\m\n} \chi^m )+\frac{i}{16} \g_{\m\n\r} \chi^m (\bar{\e} 
\g^{\n\r} \chi^m ) \nonumber \\ & & \quad \quad - \frac{9i}{16} v_r 
c^{rab} 
\l^{a} (\bar{\e} \g_\m \l^{b})+  
\frac{i}{16} v_r c^{rab} \g_{\m\n} \l^{a} (\bar{\e} \g^\n \l^{b} ) - 
\frac{i}{32}
v_r c^{rab}  \g^{\n\r} \l^{a} (\bar{\e}
\g_{\m\n\r} \l^{b} ) \quad ,\nonumber\\  
& & \delta \chi^m =
\frac{i}{2} x^m_r (\hat{\de_\m v^r} ) \g^\m \e +
\frac{i}{12} x^m_r \hat{H}^r_{\m\n\r} \g^{\m\n\r}\e +
\frac{1}{4} x^m_r c^{rab} \g_\m \l^{a} (\bar{\e} \g^\m \l^{b} )  \quad
,\nonumber\\  
& & \delta \l^{a} =-\frac{1}{2\sqrt{2}}\hat{F}^{a}_{\m\n} 
\g^{\m\n} \e  +[(v \cdot c )^{-1} (x^m \cdot c )]^{ab} \lbrace -
\frac{1}{2} (\bar{\chi}^m \l^{b} ) \e 
- \frac{1}{4} (\bar{\chi}^m \e ) \l^{b}  
\nonumber \\ 
& & \quad \quad + \frac{1}{8} 
(\bar{\chi}^m \g_{\m\n} \e ) \g^{\m\n}
\l^{b} \rbrace \nonumber
\eeq 
gives the supersymmetry anomaly
\beq  {\cal{A}}_\e & &=c_r^{ab} c^{r cd}  \lbrace -\frac{1}{16}
\e^{\m\n\a\b\g\delta}\delta_\e A^{a}_\m A^{b}_\n F^c_{\a\b} F^d_{\g\delta}
-\frac{1}{8}
\e^{\m\n\a\b\g\delta} \delta_\e A^{a}_\m F^{b}_{\n\a} 
A_{\b}^{c}F_{\g\d}^{d} \nonumber\\ & & +\frac{i e}{8} \delta_\e A^{a}_\m 
F^{b}_{\n\r}
(\bar{\l}^c
\g^{\m\n\r} 
\l^d )+\frac{i e}{8} \delta_\e A^{a}_\m (\bar{\l}^{b} \g^{\m\n\r} \l^c )
F^d_{\n\r}  + \frac{ie}{4}\delta_\e A^{a}_\m (\bar{\l}^{b}\g_\n \l^c ) 
F^{d \m\n}
\nonumber\\ & & -
\frac{ie}{128} (\bar{\e}\g^\a \psi_\m ) (\bar{\l}^{a} \g^{\m\n\r} \l^{b} )
(\bar{\l}^c
\g_{\a \n\r} \l^d )  -\frac{e}{8\sqrt{2}} \delta_\e A^{a}_\m 
(\bar{\l}^{b} \g^\m
\g^\n \g^\r 
\l^c )(\bar{\l}^d \g_\n \psi_\r ) \rbrace  \nonumber\\  
& & + c_r^{ab} [c^r (v \cdot c )^{-1} (x^m \cdot c )]^{cd}
\lbrace -\frac{ i}{4\sqrt{2}}
\delta_\e A^{a}_\m (\bar{\l}^{b} \g^\m \l^c )(\bar{\l}^d \chi^m )
\nonumber\\
& & +\frac{i}{16 \sqrt{2}}\delta_\e A^{a}_\m (\bar{\l}^b \g^\m \g^{\n\r}
\l^d )(\bar{\chi}^m \g_{\n\r}\l^c ) 
-\frac{i}{8\sqrt{2}} \delta_\e A^a_\m (\bar{\l}^b \g^\m \l^d )(\bar{\chi}^m
\l^c ) \rbrace \nonumber\\
& & +\frac{\a}{8} c_r^{ab}c^{rcd}\delta_\e \lbrace
e (\bar{\l}^a \g_\m \l^c )(\bar{\l}^b \g^\m \l^d )\rbrace \quad . 
\nonumber
\eeq 
Summarizing, the complete theory would contain additional non-local 
couplings induced by fermion loops, whose variation would cancel the 
anomalous contribution of the contact terms. Thus,  the low-energy 
couplings that we are displaying are properly 
neither gauge-invariant nor supersymmetric. 
However, gauge and supersymmetry anomalies are related by Wess-Zumino 
consistency conditions, and this grants the coherence of the construction.
The presence of the arbitrary parameter $\a$ reflects the
freedom of adding to the anomaly the variation of a local functional, 
consistently with all Wess-Zumino conditions. 
This anomalous behavior of the low-energy Lagrangian is related to
another remarkable property of these models \cite{frs}:
aside from local symmetry transformations and the equation of motion, 
the commutator of two supersymmetry transformations on the  gauginos generates
the two-cocycle
\beq
 \delta_{(\a)} \l^a & &=
[(v \cdot c )^{-1} c_r ]^{ab} c^{rcd}[-\frac{1}{8}(\bar{\e}_1
\g_\m \l^c )(\bar{\e}_2
\g_\n \l^d ) \g^{\m\n} \l^b -\frac{\a}{4} (\bar{\l}^b \g_\m
\l^c )(\bar{\e}_1 \g_\n \l^d )
\g^{\m\n} \e_2 \nonumber\\  
& & +\frac{\a}{32}(\bar{\l}^b\g_{\m\n\r}\l^c )(\bar{\e}_1 \g^\r 
\l^d ) \g^{\m\n} \e_2 +\frac{\a}{32} (\bar{\l}^b \g_\r
\l^c )(\bar{\e}_1
\g^{\m\n\r} \l^d ) \g_{\m\n} \e_2  \nonumber\\  
& & + \frac{1-\a}{8} (\bar{\l}^b \g_\m
\l^c ) (\bar{\e}_1
\g^\m \l^d ) \e_2 -(1 \leftrightarrow 2)  
+ \frac{1-\a}{32} (\bar{\e}_1 \g^\m \e_2 )(\bar{\l}^c
\g_{\m\n\r} \l^d  )\g^{\n\r}\l^b ] \quad ,\nonumber
\eeq 
different from zero for any value of $\a$.
In six dimensions the Wess-Zumino conditions
close only on the field equations of the gauginos, and
this two-cocycle actually makes these conditions close for any value 
of $\a$.
In the case of a 
single vector multiplet, in which the term of the Lagrangian 
proportional to $\l^{4}$
disappears, the two-cocycle is still present, 
although it is properly independent of $\a$.

Although the properties of (anti)self-dual tensors imply that 
the $H^2$ term in the
Lagrangian vanishes identically, the field equations  
obtained varying ${\cal{L}}$ are the ones that result from the supersymmetry 
algebra when the (anti)self duality conditions are imposed. 
It is possible to apply the construction of Pasti, Sorokin and 
Tonin (PST) \cite{pst} in this case, following the results of 
\cite{rs2}, where this construction was applied to the case of 
(anti)self-dual tensors coupled to non-abelian vector multiplets, 
thus generalizing the results of 
\cite{6bpst}, where only tensor multiplets were considered.  
Since the theory describes a single self-dual 2-form
$$
\hat{\cal{H}}_{\m\n\r} =v_r \hat{H}^r_{\m\n\r} -\frac{i}{8}(\bar{\chi}^m 
\g_{\m\n\r} \chi^m )
$$
and $n$ antiself-dual 2-forms
$$
\hat{\cal{H}}^m_{\m\n\r} =x^m_r \hat{H}^r_{\m\n\r} +\frac{i}{8}x^m_r c^{rab}
(\bar{\l}^a \g_{\m\n\r} \l^b ) \quad ,
$$
the complete Lagrangian is obtained by the addition of the term
$$
-\frac{1}{4}\frac{\de^\m \phi \de^\s \phi }{(\de \phi )^2 }
[ \hat{\cal H}^-_{\m\n\r} \hat{\cal H}^-_{\s}{}^{\n\r} +
\hat{\cal H}^{m+}_{\m\n\r} \hat{\cal H}^{m+}{}_\s{}^{\n\r} ] \quad ,
$$
where $\phi$ is an auxiliary scalar field \cite{pst,6bpst} and 
$H^{\pm}=H \pm *H $. This lagrangian has PST gauge 
invariances needed to cancel the additional degrees of 
freedom \cite{pst}.
Once the transformations of the gravitino and of the tensorinos
are properly modified,
the supersymmetry algebra generates also the PST 
gauge transformations
\cite{6bpst}.
The field equations obtained from the complete Lagrangian reduce to those
obtained from the Lagrangian without the PST term, once these gauge
invariances are fixed \cite{pst}. 
Alternative ways of obtaining a lagrangian formulation for self-dual tensors 
can be found in the literature. While in general these involve
infinite auxiliary fields \cite{infinite},  Kavalov 
and Mkrtchyan \cite{km} had obtained long ago a complete action for pure 
d=6 (1,0) supergravity in terms of a single tensor auxiliary field.
Their work may be connected to the result of \cite{6bpst} 
via an ansatz relating their tensor to the PST scalar.

As supersymmetry does not constrain the values of the 
coefficients $c^{r}$, we have obtained a class of models whose anomaly 
polynomials can contain odd powers of the individual 
field strengths $F^{a}$. It is interesting to compare these results with 
\cite{cjlp}. Although for generic values of the $c$'s
the $SO(1,n)$ global symmetry is broken, the authors of \cite{cjlp} 
consider the amusing case of $n=2$ with two abelian vector 
multiplets transforming in the spinorial representation of $SO(1,2)$.
Identifying this group with the one that transforms the tensor fields, 
one obtains an $SO(1,2)$-invariant
Lagrangian if $c^{r}=\g^{0}\g^{r}$. In particular the results of 
\cite{cjlp} correspond to the Majorana representation of $SO(1,2)$:
$$
\g^{0}=\s_{2}\quad , \qquad \g^{1}=i \s_{1} \quad , \qquad \g^{2}=i \s_{3}
\quad ,
$$
and for this choice the anomaly polynomial vanishes identically.

The transition to tensionless strings corresponds to values of the 
scalar fields for which the gauge coupling vanishes \cite{tensionless}. In our 
Lagrangian, this would correspond to the vanishing of some eigenvalues of the 
matrix $v_r c^{r ab}$. In the case of \cite{cjlp} the moduli space
is a two-dimensional hyperboloid, described by the equation $v_0^2 -v_1^2 -v_2^2
=1 $, and one can show that 
the eigenvalues of the matrix $v_r c^{r ab}$ are both positive for
$v_0 \geq 1$ and both negative for $v_0 \leq -1$, so that the transition 
is not reached. 

\vspace{1cm}

It is a pleasure to thank D. Anselmi and in particular A. Sagnotti
for useful discussions.

\end{document}